# A method to decrease the harmonic distortion in Mn-Zn ferrite/PZT and Ni-Zn ferrite/PZT layered composite rings exhibiting high magnetoelectric effects.

V. Loyau[1], V. Morin[1], J. Fortineau[2], M. LoBue[1], and F. Mazaleyrat[1]

[1]SATIE UMR 8029 CNRS, ENS Cachan, Université Paris-Saclay, 61, avenue du président Wilson, 94235 Cachan Cedex, France.

[2] INSA Centre Val de Loire, Université François Rabelais, Tours, GREMAN UMR 7347, Campus de Blois, 3 rue de la chocolaterie, CS 23410, 41034 Blois Cedex, France

*Abstract.*

We have investigated the magnetoelectric (ME) effect in layered composite rings subjected to circumferential AC magnetic fields and DC magnetic fields in radial, axial or circumferential directions. Bilayer samples were obtained combining different grades of commercial Mn-Zn ferrites or Ni-Zn ferrites with commercial lead zirconate titanate (PZT). Mn-Zn ferrites with low magnetostriction saturation ($\lambda_s < 10^{-6}$) and low magneto-crystalline anisotropy constants show high ME capabilities when associated with PZT in ring structures. In certain conditions, these ME effects are higher than those obtained with Terfenol-D/PZT composites in the same layered ring structure. Magnetostrictive and mechanical characterizations have given results that explain these high ME performances. Nevertheless, Mn-Zn ferrite/PZT composites exhibit voltages responses with low linearity especially at high signal level. Based on the particular structure of the ME device, a method to decrease the nonlinear harmonic distortion of the ME voltages is proposed. Harmonic distortion analysis of ME voltages measured in different configurations allows us to explain the phenomenon.

## I. INTRODUCTION

In the field of magnetoelectric (ME) materials, ME composites consisting in piezoelectric and magnetostrictive phases mechanically coupled to each other have shown a great interest. The wide range of piezoelectric and magnetostrictive materials permits to obtain a large variety of ME composites with different properties and application areas. Field and current sensors[1,2,3,4] (direct ME effect), ME memories[5,6], and electrostatically tunable inductances[7,8] (converse ME effect), are promising fields for ME composites. Concerning the current sensors, ME composites are well suited for manufacturing devices with low power consumption and simplified electronic. Among the different possible connectivities between the two phases, only two have shown great interests: 3-3 (bulk composites[9,10,11]), and 2-2 (laminated composites[12,13,14]). When fabricated, ME composites are characterized using a standardized set-up[15]: in case of direct ME effect measurements, ME samples are subjected to unidirectional AC and DC magnetic fields and the induced electric field across the piezoelectric layer is sensed. Obviously, ME samples optimized with respect to this method of measurement have high sensing capabilities when a AC field with a radial direction is produced by an external coil carrying AC currents. However, for current sensors there is a need to measure currents flowing in a straight wire, producing circumferential magnetic fields. In this case, the ME sample best geometry is the ring configuration. In the literature, there are few works reported on the subject of ME composites in such a structure. Dong et *al.* (see Ref. 1 and 2) have proposed ME rings in laminate configurations consisting in a PZT or PZN-PT ring sandwiched between two Terfenol-D rings. This configuration was named C-C coupling mode because the Terfenol-D layers was magnetized in the circumferential direction and the piezoelectric layer was polarized in this same direction. Magnetic bias was produced by external permanent magnets. More recently, Leung et *al.* (see Ref. 3) have tailored a smart ME ring. The structure is based on two concentric layers. The external one is a bulk PZT ring bonded to an inner ring consisting in a Terfenol-D/Epoxy-matrix composite. Small NdFeB permanent magnets are included in the magnetostrictive phase for magnetic biasing. The PZT ring was polarized in the radial direction, whereas the magnetic field is circumferential.

In a recent study[4], we reported that optimized compositions of Co-substituted Ni-Zn ferrites made by SPS (Spark Plasma Sintering) produce high ME effect in transversal mode when associated with PZT in layered composites. The ME performances are comparable to those obtained with Terfenol-D/PZT composites in the same structure. We demonstrated that the low piezomagnetic coefficients of the ferrite material (in comparison with Terfenol-D) are counter balanced by a high stiffness that permits to produce high mechanical stress. Secondly, we have shown that the permeability of the magnetic layers, associated with the demagnetizing factor, influences the ME responses. Using a new method of ME characterization, based on ME samples in a ring structure submitted to a circumferential AC field, the demagnetizing effect was overcame, and the intrinsic ME behavior was measured. It was established that low susceptibilities ($\chi \leq 200$) (so high field penetrations) are needed to obtain high ME effects when the AC magnetic field is produced by an external means (Helmholtz coils for example). In this case, Terfenol-D and nano-size grains Ni-Zn ferrites produce the best ME effect. But when a circumferential AC field is forced in a ring structure, it was shown that the ME response is not weakened by a high susceptibility, because the demagnetizing effect does not exist.

Starting from the previous considerations, we have shown, in the present paper, the potential interest of ordinary high permeability Mn-Zn ferrites and Ni-Zn ferrites in ME layered composite ring structures subjected to circumferential AC magnetic field. In this work, we have investigated ME effects in bilayered ME rings consisting in PZT rings pasted on commercial high permeability Mn-Zn ferrites and Ni-Zn ferrites rings. Those commercial ferrites are commonly used to make inductors, transformers, and filters. To our best knowledge, those kinds of magnetic materials were never used in ME applications. The ME characterization set-up involves the use of circumferential AC field and radial, axial or circumferential DC magnetic fields. Since the AC magnetic field is not subjected to demagnetizing effect, the theory predict that the ME response is not weakened even for magnetic materials exhibiting high permeability such as Mn-Zn ferrites. For comparison, bilayered ME rings of the same structure made with high performance Terfenol-D or Ni-Co-Zn ferrite (with nano-size grains) were characterized. Then, the different ME samples were tested in current sensor configurations and their potential uses were discussed. We have focused our work on a study of the ME voltage harmonic distortion. The level of fundamental, second and third harmonics of ME voltages were analyzed for AC magnetic fields increasing from 10 A/m up to 2000 A/m. A special configuration of electrodes and electric field measurement set-up across the piezoelectric layer have permitted to develop an original method to decrease the harmonic distortion of the ME voltage in ring structures. Lastly, exploiting the low magneto-crystalline anisotropies of Mn-Zn ferrites, we proposed a special configuration of ME current sensors. The bias magnetic field was produced without external permanent magnets or external electromagnets. This is a promising way to reduce the size of ME sensors.

The paper is organized in the following way. In Sec. II, a theoretical analysis of the ME effects, and the influence of the demagnetizing field in two different configurations of ME measurement is discussed. In Sec. III, the fabrication method of the ME samples is presented, and the ME characterization set-up is described. Characterization results are given in Sec IV. Firstly, ME measurements for small AC signal show that all samples produce ME voltages in the same range. Magnetostrictive and mechanical characterizations were conducted on a Mn-Zn ferrite ring. The measured parameters were used in a theoretical model of the ME coefficient that fit well the experimental. In section V, a study at higher levels of AC signal demonstrates that Mn-Zn ferrites have piezomagnetic effects exhibiting low linearity in comparison with Terfenol-D or Ni-Co-Zn ferrite. In order to overcome this problem, we have proposed a differential voltage measurement method that increase the linearity of the ME response. A harmonic analysis of the distortion shows that this method permits to weaken the second harmonic components and it leads to a decrease of the global distortion of the ME signal. Lastly, we have studied a current sensor structure where the DC bias field was produced without permanent magnets.

## II. THEORETICAL ANALYSIS

Usually, ME composites are characterized by applying a small external AC magnetic field (by means of Helmholtz coils) superimposed to a DC field[15] in the same direction. The external AC field is measured by means of a hall probe or a search coil. When a DC bias field and a AC field $H_1^a$ are applied in direction (1), producing an electric field $E_3$ in the direction (3), the transversal coupling ME coefficient $\alpha_{31}$ (in quasi-static mode) is theoretically given by[4]:

$$\alpha_{31} = \frac{E_3}{H_1^a} = -\frac{d_{31}^e}{\varepsilon_{33}^T \left[s_{11}^E + s_{12}^E + \gamma(s_{11}^H + s_{12}^H)\right] - 2(d_{31}^e)^2} \times \frac{(d_{12}^m + d_{11}^m)}{1 + N_r \chi} \quad (1)$$

where $s_{11}^E$ and $s_{12}^E$, $d_{31}^e$, and $\varepsilon_{33}^T$, are zero field compliances, piezoelectric coefficient, and zero stress permittivity, respectively, for the piezoelectric material; $s_{11}^H$ and $s_{12}^H$, $d_{12}^m$ and $d_{11}^m$, and $\chi$ are zero field compliances, intrinsic piezomagnetic coefficients, and zero stress dynamic susceptibility, respectively, for the magnetic material; $\gamma$ and $N_r$ are the volume ratio of the materials and the radial demagnetizing factor respectively. Note that for most of polycrystalline materials, $(d_{12}^m + d_{11}^m) \sim d_{11}^m/2$. Furthermore, it must be noted that a perfect (but not realistic) mechanical coupling between the piezoelectric and piezomagnetic layers is considered here (without any flexural strain). Due to the demagnetizing effect, the level of the internal AC field is divided by $(1 + N_r \chi)$ with respect to the level of the external AC field. The left hand term of Eq. 1, depending on the piezoelectric properties and the mechanical properties of the magnetic and piezoelectric materials, can be regarded as constants. The right hand term depends strongly on the bias field because the values of $d_{12}^m$, $d_{11}^m$, and $\chi$ are dependent on the field. As a consequence, the ME curve is shaped by the right hand term.

A second method was recently proposed[4] to characterized ME samples. The aim was to overcome the demagnetizing effect and reach the intrinsic ME behavior. A circumferential AC magnetic field is forced within the magnetic material by means of a coil wounded on the ME ring (see Fig. 1(a)). In this configuration, the value of the internal AC field $H_{1,2}$ is deduced from the measurement of the current flowing into the coil. Obviously, in this case, the demagnetizing factor $N_r$ in Eq. 1 is equal to zero. This leads to the transversal coupling ME coefficient $\alpha_{31,2}$ related to the circumferential AC field $H_{1,2}$ in the (1,2) plane:

$$\alpha_{31,2} = \frac{E_3}{H_{1,2}} = -\frac{d_{31}^e}{\varepsilon_{33}^T \left[s_{11}^E + s_{12}^E + \gamma(s_{11}^H + s_{12}^H)\right] - 2(d_{31}^e)^2} \times d_{11}^m/2 \qquad (2)$$

As seen in Eq. 2, the dynamic susceptibility $\chi$ has no influence on the amplitude of the ME coefficient $\alpha_{31,2}$. Nevertheless, the piezomagnetic coefficient, $d_{11}^m$, depends strongly on the internal bias field $H_{DC}$. In general, an applied bias field $H_{DC}^a$ is produced by an electromagnet and is measured by means of a Hall probe. Then, the ME coefficient $\alpha_{31,2}$ is plotted as a function of the applied bias $H_{DC}^a$. However, this bias field is subjected to the demagnetizing effect and the link between internal and applied fields is given by:

$$H_{DC} = \frac{1}{1+N_r \cdot \chi_{DC}} H_{DC}^a \qquad (3)$$

where $\chi_{DC}$ is the static susceptibility. Thus, the demagnetizing effect has influence on the shape of the ME coefficient curve, shifting it along the $H_{DC}$ axis.

When a AC field is applied by external sources (Helmholtz coils, for example), the demagnetizing effect reduces the field penetration within the magnetic material. Eq. 1 clearly demonstrates that, in this condition of measurement, magnetostrictive materials with low susceptibilities $\chi$ produce high ME effects when associated with PZT. For example, in Ref. 12, the transversal ME coefficient was 0.8V/A for Ni-Zn ferrite/PZT multilayers with optimized compositions. In Ref. 16, similar compositions of ME composites gave $\alpha_{31} = 0.525\ V/A$ (trilayer sample) and $\alpha_{31} = 0.195\ V/A$ (bilayer sample). The differences in ME coefficients seem to be due to demagnetizing effects: the ferrite layers have thicknesses with different values. Usually, $\chi \sim 50 - 100$ for polycrystalline Ni-Zn ferrites with nano-sized grains, and $\chi \sim 10 - 30$ for Terfenol-D (at optimal bias). It is obvious that in this case, materials with high susceptibilities such as Mn-Zn ferrites or Ni-Zn ferrites (with grain size over 5µm) cannot show exploitable ME effects. By opposition, when a circumferential field $H_{1,2}$ is forced within the ferrite, these materials are not limited by their high susceptibilities (see Eq. 2). As a consequence, the only condition that requires a high ME effect is a high intrinsic piezomagnetic coefficient $d_{11}^m = (d\lambda_1/dH_1)_{H_{DC}}$ (where $\lambda_1$ is the magnetostriction measured is direction (1) for an internal field $H_1$). Knowing the saturation magnetostriction $\lambda_s$ and the internal field $H_s$ at this saturated state, this coefficient can be estimated roughly by applying the formula: $d_{11}^m \sim \lambda_s/H_s$. In a simple model[17], $H_s$ is directly linked to $K_1$, the magneto-crystalline anisotropy constant of the second order, and we can conclude that the intrinsic piezomagnetic coefficient $d_{11}^m$ is proportional to the ratio $\lambda_s/K_1$. The magnetostriction and the magnetic anisotropy have the same origin, namely the spin-orbit coupling. Ferrite properties can be tailored by the chemical composition, so $\lambda_s$ and $K_1$ can vary on a large span. But, in general, magnetostriction and magneto-crystalline anisotropy change in the same manner (i.e. low magnetostriction is coupled with low anisotropy, whereas high magnetostriction is coupled with high anisotropy). We can expect that almost all ferrite compositions (Mn-Zn ferrites, Ni-Zn ferrites, Co-Zn ferrites and substituted compositions) exhibit $\lambda_s/K_1$ ratios (and so, intrinsic piezomagnetic coefficients $d_{11}^m$) within the same order of magnitude. Optimized compositions of Ni-Zn ferrites have much lower piezomagnetic coefficients in comparison with Terfenol-D. But it was demonstrated that this weakness is counter balanced by a low compliance, so high ME effects were obtained in layered ME composites[4]. Comparable phenomenon can be expected with Mn-Zn ferrites in layered ME structures, even for compositions exhibiting very low saturation magnetostriction.

## III. SAMPLES FABRICATION AND MEASUREMENT SET-UP

Commercial ferrite rings were purchased from Ferroxcube: two Mn-Zn ferrites exhibiting low magnetostriction, 3E6, and 3E8, and a Ni-Zn ferrite, 4A11. All of them were machined to reduce their thicknesses to 2 mm. To form ME bilayer samples, each ferrite ring was pasted with silver epoxy (Epotek E4110) on a PZT ring (PIC255, 10 mm outer diameter, 5 mm internal diameter, 1 mm thickness, polarized in direction (3)). For comparison, two ME bilayer samples were fabricated using the same technique with materials known to produce high ME effect: a Terfenol-D (TFD) ring purchased from Etrema, and a Ni-Zn ferrite (FNCZ) ring made by reactive Spark Plasma Sintering (SPS) in our laboratory with the initial composition (mixture of precursor oxides) $(Ni_{0.973}Co_{0.027})_{0.75}Zn_{0.25}Fe_2O_4$ (see Ref. 4 for information on the synthesis process). The final composition after the SPS stage was measured using EDX chemical analysis (Hitachi S-3400N SEM) and it is very close to the initial one. This is due to the very short time duration (25 minutes) of the SPS stage. This composition of ferrite was chosen because it produces high ME effects when a circumferential AC magnetic field is applied. The characteristics of all the magnetic material are summarized in Table 1. Lastly, five bilayered ME samples were fabricated: 3E6/PIC255, 3E8/PIC255, 4A11/PIC255, FNCZ/PIC255, and TFD/PIC255.

The ME characterization procedure is based on an 8 turns coil wounded on the ME ring that forces a circumferential AC magnetic field within the magnetic layer. The internal magnetic field $H_{1,2}$ in the (1,2) plane is related to the AC current $I_{AC}$ flowing into the coil: $H_{1,2} = n.I_{AC}/L$, where $n = 8$ is the number of turns and $L$ is the mean length of the circular magnetic path. When the DC magnetic field is applied in the (radial) direction (1) by the means of an electromagnet (see Fig. 1(a)), the measurement configuration is named CRA (Circumferential AC field, Radial DC field, Axial electric field). In this case, two opposite voltages $V$ and $V'$ are induced on each halves electrodes on the top of the PZT layer (the top electrode was split by two strokes of file in the same direction (1)). When the DC magnetic field is applied in the (axial) direction (3) (see Fig. 1(b)), the measurement configuration is named CAA (Circumferential AC field, Axial DC field, Axial electric field). Lastly, when the AC and DC magnetic fields are forced in the circumferential direction (see Fig. 1(c)), the measurement configuration is named CCA (Circumferential AC field, Circumferential DC field, Axial electric field). It must be noted that the electric field is always in the (axial) direction (3). A lock-in amplifier measures the voltage $V$ on a half electrode at each DC magnetic field. For all those working points, the ME coefficient is deduced from $\alpha_{31,2} = V/t/H_{1,2}$, where $t = 1\ mm$ is the thickness of the PZT layer, and $H_{1,2}$ is the internal AC field deduced from the measurement of the current $I_{AC}$ flowing into the coil.

## IV. ME CHARACTERIZATION.

### A. ME coefficient measurements in CRA configuration.

The ME coefficients in CRA configuration were measured and plotted in Fig. 2 for the ME samples made with ferrites and in Fig. 3 for the ME sample made with Terfenol-D. The magnitude of the AC magnetic field is around 1.5 A/m at 80 Hz (to avoid any resonance phenomenon). Concerning the ferrite/PIC255 samples, the maximum ME effects occur at low bias magnetic fields between 8 kA/m and 25 kA/m. This fact confirms that the ferrite materials (due to high permeability) are easily magnetized in comparison to the Terfenol-D: the TFD/PIC255 sample exhibit a maximum ME effect at $H_{DC}^a = 90\ kA/m$. On the other hand, it is clearly seen that all the ferrite/PIC255 samples have better ME effects ($0.76\ V/A < \alpha_{31} < 3.3\ V/A$) than the ME sample fabricated with Terfenol-D ($\alpha_{31,2} = 0.7\ V/A$). Note that the ME curve of the 3E6/PIC255 sample exhibits two peaks: a small peak at low DC field ($H_{DC}^a \sim 6\ kA/m$) and a bigger one at higher DC field ($H_{DC}^a \sim 20\ kA/m$). The reason of this behavior will be explained later in the paper. The high ME effect obtained with Mn-Zn ferrites is not obvious and it can be explained by low compliances (two times lower than the PZT compliances, see Table 2) associated to relatively high intrinsic piezomagnetic coefficients, according Eq. 2.

### B. ME coefficient measurements in CAA configuration.

A second configuration to measure a ME coupling is investigated: the DC magnetic field is applied in the direction (3) when the circumferential AC magnetic field is in the (1,2) plane of the ME sample. Results are plotted in Fig. 4 for 3E8/PIC255 and TFD/PIC255 ME samples. The two curves show ME effects roughly 10 times lower in comparison to the previous configuration measurement. Since the demagnetizing effect do not exist for the AC magnetic field, this discrepancy of the ME effects can be explained by a lowering of the piezomagnetic coefficients when the applied DC field is perpendicular to the AC field. In fact, when a bias field is applied in direction (3), the magnetization is roughly forced in this direction. So, the AC field, which is applied in a perpendicular direction, has difficulties to turn the magnetization, and consequently the piezomagnetic effect is low. Moreover, due to higher demagnetizing effects (which exist for the DC magnetic field) peaks of curves are shifted at higher bias fields. Note that for the TFD/ PIC255 sample there are two peaks at $H_{DC}^a = 220\ kA/m$ and $650\ kA/m$.

### C. Correlation with magnetostriction measurement on 3E6 material.

High permeability Mn-Zn ferrites can produce ME effects (when associated with PZT) comparable to those obtained with Terfenol-D. This good performance should be explained by measuring the piezomagnetic and elastic properties. It is known that high susceptibility Mn-Zn ferrites exhibit low magnetostriction, in the range of $10^{-6}$ (in relative) at saturation. This value is too low to be measured by means of metal strain gauge, and piezoresistive gauges are very sensitive to thermal drift. So we have chosen an interferometry method that permits strain measurement under $10^{-8}$. Magnetostriction measurement was conducted on a 3E6 ring. Four turns of wire were wounded on the ring and a current (1.2 A peak) at 1kHz was applied. This current forces a 200A/m peak magnetic field within the ring, which is sufficient to reach the saturation of the magnetic material. The velocity, at the surface of the ring along a radial direction, was measured by means of a velocimeter (Polytec Laser Surface Velocimeter). The displacement, and then the strain, was deduced by an integration of the velocity. Magnetostriction curve versus internal magnetic field is plotted in Fig. 5. A well known butterfly shape is obtained. Note that the velocimeter was used at its lowest frequency limit (1kHz) and the curve may be a little bit distorted. The magnetostriction is very low, $\lambda_s < 0.3\ ppm$, but it is obtained for a very low internal field: $H \sim 200\ A/m$. We can estimate an intrinsic piezomagnetic coefficient: $d_{11}^m = (d\lambda/dH)_{H_{DC}} \sim 3 \times 10^{-9} m/A$ at the optimal bias $H_{DC} \sim 80\ A/m$. Velocity wave measurement of ultrasonic pulses (20MHz center frequency) along the thickness of the ring permits us to deduce the $s_{33}^H$ and $s_{31}^H$ compliances of the 3E6 material. The pulse-echo technique for longitudinal and shear waves leads to: $V_l = 6093 m/s$ (longitudinal) and $V_t = 3429 m/s$ (transversal). This yields to: $s_{33}^H = 7 \times 10^{-12} m^2/N$ and $s_{31}^H = -1.9 \times 10^{-12} m^2/N$. If we assume that the polycrystalline ferrite is not textured, we have: $s_{11}^H = s_{33}^H$ and $s_{12}^H = s_{31}^H$. Using Eq. 2 and the data given in Table 2, we have calculated the magnetoelectric response of the sample # 3E6/PIC255. We obtain theoretically: $\alpha_{31,2} \sim 1.27\ V/A$, and experimentally the value is $\alpha_{31,2} \sim 0.9\ V/A$ (at optimal bias), which is lower. Eq. 2 has been developed assuming that the strain in the magnetic layer, $S_1^m$, and the strain in the piezoelectric layers, $S_1^e$, are the same. That is definitely a rough approximation. To try to better describe the reel system we introduce a ratio of strain $\eta$, taking into account the differential strain between the two layers: $\eta = \langle S_1^e \rangle / \langle S_1^m \rangle$ (in average). So Eq. 2 becomes[4]:

$$\alpha_{31,2} = \frac{E_3}{H_{1,2}} = -\frac{\eta\ d_{31}^e}{\varepsilon_{33}^T [s_{11}^E + s_{12}^E + \eta\gamma(s_{11}^H + s_{12}^H)] - 2(d_{31}^e)^2} \times d_{11}^m/2 \qquad (4)$$

The ratio $\eta$ depends on the composite structure. The lower value is obtained for staked PZT/ferrite bilayer structures when the PZT layer is stressed on only one face. When the PZT layer is stressed on its both faces (in case of ferrite/PZT/ferrite trilayers or more) the value of $\eta$ is increased[4]. Moreover, in case of concentric PZT/ferrite discs[19], the mechanical coupling is enhanced. In the present case of a staked bilayer ring, the hole can influence the mechanical coupling between the layers, and consequently, the value of $\eta$ is affected. So the influence on the ME response of the composite ring structure (that affects the mechanical coupling) is included in $\eta$. It is difficult to measure or to calculate the value of the ratio of strain $\eta$. Nevertheless, using $\eta = 0.65$ in Eq. 4, the theoretical value of the ME coefficient, $\alpha_{31,2}$, fits the experimental one. This value of $\eta$ is in the range of a value already measured for a ME bilayer of the same kind[4]. Note that the magnetostriction curve given in Fig. 5 shows two local extrema at $H \sim 30\ A/m$ (local maximum) and $H \sim 50\ A/m$ (local minimum). It may be due to a special shape of the domain structure near the demagnetized state. This local behavior at low field can explain the first small peak of the ME curve (see Fig. 2) for the 3E6/PIC255 sample. The previous magnetostriction loop measurement method developed for the 3E6 ring cannot be applied to Ni-Zn ferrites (4A11 and FNCZ) or Terfenol-D rings. In fact, due to their high magneto-crystalline anisotropies, very high level currents ($I > 30A$) in the coils are needed to saturate the materials. Nevertheless, the intrinsic piezomagnetic coefficient can be roughly estimated using the approximated formula: $d_{11}^m \sim \lambda_s/H_s$, where $\lambda_s$ is the saturation magnetostriction, and $H_s$ is the internal field needed to saturate the material. The saturation magnetostriction values were measured on disc samples (see method in Ref. 4) using strain gauges and we obtained: $\lambda_s = 19\ ppm$ (FNCZ) and $\lambda_s = 1000\ ppm$ (Terfenol-D). The corresponding internal saturation fields were deduced from virgin magnetization curves (corrected in demagnetizing fields) measured by means of a Vibrating Sample Magnetometer (Lakeshore 7404). We obtained: $H_s = 2.5 \times 10^4\ A/m$ (FNCZ) and $H_s = 10^5\ A/m$ (Terfenol-D). So, the estimated intrinsic piezomagnetic coefficients are: $d_{11}^m \sim 0.8 \times 10^{-9}\ m/A$ (FNCZ), and $d_{11}^m \sim 10 \times 10^{-9}\ m/A$ (Terfenol-D). Using the same rough calculation method for the 3E6 material, we obtained: $d_{11}^m \sim 0.2 ppm/\ 200 = 10^{-9} m/A$ (see Fig. 5 for $\lambda_s$ and $H_s$ values). The approximated piezomagnetic coefficients of 3E6 and FNCZ materials are in the same range. This fact is confirmed by results in Fig. 2: the two materials produce ME responses with the same level. Although the Terfenol-D material have a piezomagnetic coefficient roughly ten times higher in comparison to the previous materials, the ME responses is in the same range ($\alpha \sim 0.7\ V/A$). In the case of Terfenol-D, the high piezomagnetic coefficient is counter balanced by its high compliance. (see Ref. 4).

## V. STUDY OF CURRENT SENSOR APPLICATIONS.

### A. ME samples used in CRA configuration.

We have investigated the potential use of those ferrite/PZT bilayer ME samples in a current probe application. The set-up is described in Fig. 1(a), where the DC field was applied along the direction (1) and the AC field in the (1,2) plane was produced by a 8 turns coil wounded on the ME ring (CRA configuration). A 55 mA peak, triangle waveform current at 1kHz was applied to the coil. It correspond to a medium level AC field of about 18A/m. In each case, the DC field was set to obtain the highest ME voltage with good linearity. The time variation of the applied current was deduced from the voltage produced by a 1Ω resistor connected in series with the coil. The ME voltage was sensed directly by a passive 1 ÷ 10 voltage probe (10 MΩ input impedance) and recorded by an oscilloscope (Lecroy Waverunner 44Xi). The results are given in Fig. 6. All the ME samples show good linearity, except the FNCZ sample. For this medium level AC field, the voltage levels produce by the ME samples are consistent with the ME coefficient measurements (Fig. 2 and Fig. 3), except for the 4A11/PIC255 sample, for which the voltage at medium excitation is two times higher with respect to the ME coefficient measurement. Note that as expected, the TFD/PIC255 sample shows the lowest voltage response (but with the best linearity).

For AC field over 50 A/m, high distortions on ME voltages appear for 3E6/PIC255, 3E8/PIC255, and 4A11/PIC255 samples. It means that these samples are suitable only for low AC magnetic field (or low AC current) detection. On the other hand, ME samples made with TFD or FNCZ materials show relevant ME coefficients in large bias field ranges (see Fig. 2 and Fig. 3). In these cases, we can expect good linearity even when high AC magnetic fields are applied. Experiments were conducted on ME composites made with TFD and FNCZ materials under large excitation: a 3 A peak, triangle waveform current at 1kHz was applied in the 8 turns coils (the AC field is about 1200 A/m peak). The ME waveforms are given in Fig. 7. In each case, the bias field was tuned with the goal of getting a good linearity of the ME voltage. This is obtained when the bias field is over the DC field giving the maximum ME voltage. But, it conducts to a significant decrease of the ME response level. This is especially true concerning the FNCZ/PIC255 sample for which the ME coefficient is reduced to $\alpha_{31,2} = 0.26\ V/A$ at $H_{DC}^a = 35\ kA/m$ (three times lower than the low signal ME coefficient at optimal bias). This effect is less pronounced for the TFD/PIC255 sample for which the ME coefficient at high level signal, $\alpha_{31,2} = 0.52\ V/A$ at $H_{DC}^a = 200\ kA/m$, is closer to the value obtained at low AC level in Fig. 3.

When a ferrite material is subjected at the same time to a circumferential AC magnetic field and an unidirectional bias field (see Fig. 1(a)), two opposite strains occur in each half part of the material because the radial components of the AC field are opposite for radially opposed points. This effect produces two opposite electric fields in the corresponding parts of the piezoelectric layer. Theoretically, the ME coefficient can be doubled when the voltage is measured between the two halves electrodes on the top of the piezoelectric layer. Figure 8 shows an example of the two voltages *V* and *V'* sensed on each part of the FNCZ/PIC255 sample. A 300 mA peak current (1kHz) with triangular waveform was applied to the coil. The two voltages waveforms (dotted and dashed lines), measured by two passive voltage probes, are almost symmetrical: they have opposite phases, but the voltage levels are slightly different. This difference can be explained if we suppose that the piezomagnetic and the piezoelectric properties are not perfectly homogeneous in the materials. The differential ME voltage measured with only one voltage probe connected between the two halves electrodes is given in Fig. 8 (thick solid line). As expected, the voltage is twice higher and it is seen that the linearity is enhanced.

**B. Harmonic distortion analysis in CRA configuration.**

Nonlinear effect is a recent subject of study in the field of ME devices, and some applications have been developped[20, 21, 22]. As seen before in the present paper, the linearity of the ME response is enhanced when the voltage is measured between the two halves electrodes. To understand this effect, we have analyzed and compared the harmonic contents of the ME responses when the voltages are measured on a halves electrode (direct voltage) and between the two halves electrodes (differential voltage). Experiments were performed using sinusoidal AC magnetic field from 10 A/m up to 2000 A/m at 1kHz. The ME voltages were recorded by an oscilloscope (and a passive 1÷10 voltage probe) and a Fast Fourier Transform (FFT) was performed. In Fig. 9, fundamentals (1kHz), second harmonics (2kHz), and third harmonics (3kHz) are plotted as function of the AC field for the ME sample # 3E8/PIC255. It is worth noting that the DC bias field was set to obtain (relatively) low distortion at high AC field. So, the ME sensitivity is lower than the best value for this sample. Fig. 9 shows that when a differential voltage measurement is done, the fundamental and the third harmonic amplitudes are roughly doubled (+6dB) with respect to the direct voltage measurement. In an opposite way, the second harmonic is highly weakened (-18dB at $H_{ac} = 250$A/m), leading to a decrease of the global distortion of the signal. The same experiment was conducted on the sample # FNCZ/PIC255 and results are displayed in Fig. 10: here again, the differential voltage measurement leads to a decrease of the second harmonic amplitude, and this effect occurs for all even harmonics. The second harmonic compensation indicates that the second harmonic voltages on each halves electrodes are in phase, whereas the third harmonic voltages are in opposition. The second harmonic generation is mainly due to the quadratic component of the magnetostriction λ versus $H$ field behavior, $\lambda(H^2)$. In this case, the two halves parts of the magnetic material are alternatively strained in phase with same amplitude. Assuming that the related voltages are well balanced, the second harmonic component can theoretically be cancelled. Lastly, the same harmonic distortion measurements were made for the sample # TFD/PIC255, and results are displayed in Fig. 11. The bias field was chosen to produce the highest ME voltage and the lowest distortion. It can be seen that the fundamental component of the voltage increases linearly with $H_{ac}$ and the second harmonic is always 50dB lower. But for this sample, the effect of second harmonic compensation does not occur and the reason remains unclear. The third harmonic amplitude is roughly constant (~-60dBm) from $H_{ac} = 3$A/m up to 100A/m and, consequently, at low AC field ($H_{ac} < 10$A/m) the ME voltage is highly distorted. For this sample, it is clearly seen that the best ME performance are obtained when $H_{ac}$ is over 100 A/m because in this region, the harmonic distortion is low (the ME coefficient is 1.4 V/A in the differential mode). Due to the second harmonic compensation effect, the sample # FNCZ/PIC255, displays low distortion for $H_{ac}$ between 40 A/m and 200 A/m (in the differential mode). This range defines its best working region for which the ME coefficient is around 1.75 V/A. These findings demonstrate that, in a useful range of $H_{ac}$, nickel-cobalt-zinc ferrites can have performances comparable to those obtained with Terfenol-D.

**C. ME sample used in CCA configuration.**

In most of cases, ME devices need a DC magnetic bias field to reach the optimal working point (where the ME coefficient is maximum). Bias fields can be produced by permanent magnets[4] (barium ferrite magnets for example) but this technique leads to some limitations: (i) permanent magnets increase the size of the ME devices; (ii) magnetic leakage can influence other devices in the surrounding environment, (iii) the working point is not tunable. But from Fig. 6, it is apparent that Mn-Zn ferrites reach optimal piezomagnetic coefficients at low internal magnetic field (in the range of 50 A/m). This low bias field can easily be produced by an additional coil wounded on the ME ring and carrying a DC current. So we have conducted the following experiment on the 3E8/PIC255 sample. An additional coil ($N = 30$ turns) was wounded on the ME sample, carrying a DC current provided by a tunable current source (with high input impedance) made using a LM317 integrated circuit. This configuration is sketched Fig. 1(c) (CCA configuration). The $I_{DC}$ current can be set from 0 to 0.5A. The AC magnetic field was produced by a 50mV$_{rms}$, 1kHz, sinus waveform current flowing inside a 8 turns coil wounded on the ME ring. The ME voltage was measured using an oscilloscope, and has been analyzed by FFT. In Fig. 12, fundamental and second-harmonic components have been plotted as function of the DC current×turns product. Seeing the curve of the fundamental component, we find that the maximum voltage is obtained for a DC current as low as $I_{DC} = 22 mA$ ($H_{DC} = 30\ A/m$). At this working point, the ME coefficient reaches $\alpha \sim 5\ V/A$. This high ME effect can be explained by an ideal configuration where both the AC and DC magnetic fields are circumferential, and so parallel to each other. Obviously, the second-harmonic component is maximum at zero bias field. It is attributed to the frequency doubling effect due to the quadratic behavior of the magnetostriction at low field. Increasing the DC field, the second-harmonic amplitude decreases and reaches a minimum near the optimal working point. In conclusion, the feasibility of a highly sensitive tunable ME device, without permanent magnets, was demonstrated.

## VI. CONCLUSION

We have demonstrated that commercial Mn-Zn ferrites exhibiting low magnetostriction have potential interest in ME applications. A layered ring configuration with a circumferential magnetic excitation is needed to meet high ME effects. The reason for this performance is a combination of three factors: (i) the forced AC magnetic field is free of demagnetizing effect; (ii) the intrinsic piezomagnetic coefficient of the Mn-Zn ferrites is relatively high; and (iii) the compliance of ferrites is low. The present study concerning Mn-Zn ferrites in ME composites is not exhaustive. Such ferrites are widely used in electronic devices and a large amount of grades are available with various properties. We expect that various compositions with high intrinsic piezomagnetic properties could produce ME effects of potential interest. The ME ring structure is suitable for current sensing in straight cables, and it was shown that Mn-Zn ferrites are the best candidates when we need to sense low level signals. At high level signals, the non linear harmonic distortion limits the performances of the ME device. This problem is partially overcome when using the differential voltage measurement method proposed in the paper. Harmonic distortion analysis have shown that a second harmonic compensation occurs.

| Material # | Outer diameter (mm) | Internal diameter (mm) | Initial permeability (in relative) | Material type |
|---|---|---|---|---|
| 3E6 | 9.5 | 4.8 | 12000 | Mn-Zn ferrite |
| 3E8 | 9.5 | 4.8 | 18000 | Mn-Zn ferrite |
| 4A11 | 10 | 6 | 850 | Ni-Zn ferrite |
| FNCZ | 10 | 4 | 400 | Ni-Co-Zn ferrite |
| TFD | 10 | 4 | 40 | Terfenol-D |

TABLE 1. Characteristics of the magnetic rings. Characteristics of 3E6, 3E8, and 4A11 materials are cited from Ferroxcube. All the magnetic rings have 2 mm thickness.

|  | $d_{31}^e$ (pC/N) | $d_{11}^m$ (nm/A) | $s_{11}^E$ or $s_{11}^H$ (m²/N) | $s_{12}^E$ or $s_{12}^H$ (m²/N) | $\mu^T$ or $\varepsilon_{33}^T$ (in relative) |
|---|---|---|---|---|---|
| Pic255 | -180 |  | $15 \times 10^{-12}$ | $-6.5 \times 10^{-12}$ | 2400 |
| 3E6 ferrite |  | 3 | $7 \times 10^{-12}$ | $-1.9 \times 10^{-12}$ | 12000 |

TABLE 2 : Material properties for Pic255 (cited from Physik Intrumente[18]), and 3E6 ferrite.

**FIGURE CAPTIONS**

FIG.1. Sketches of three configurations of ME measurement. (a): the circumferential AC magnetic field is forced in the (1,2) plane (the coil is not represented) and the DC magnetic field is applied in the (radial) direction (1) (CRA). (b): the circumferential AC magnetic field is forced in the (1,2) plane and the DC magnetic field is applied in the (axial) direction (3) (CAA). (c): both AC and DC magnetic fields are forced in the circumferential direction (CCA). In all case, the electric field is measured in the (axial) direction (3).

FIG. 2. Transversal magnetoelectric coefficients under radial DC field (CRA configuration). Dashed line: 3E8/PIC255 sample. Dotted line: 3E6/PIC255 sample. Solid line: 4A11/PIC255 sample. Dashed-dotted line: FerriteNiCoZn/PIC255.

FIG. 3. Transversal magnetoelectric coefficient under radial DC field (CRA configuration) for the TFD/Pic255 sample.

FIG. 4. Magnetoelectric coefficients when the DC field is applied in (axial) direction (3) and the AC field is in the (1,2) plane (CAA configuration). Dashed line: 3E8/Pic255 sample. Solid line: TFD/Pic255 sample.

FIG. 5. Longitudinal magnetostriction curve versus internal magnetic field measured on a 3E6 ring excited at 1kHz.

FIG. 6. ME voltages when a 55mA peak triangle current flows in the 8 turns coil (CRA configuration). Red line: TFD/PIC255 sample. Black line: FNCZ/PIC255 sample. Green line: 3E6/PIC255 sample. Blue line: 3E8/PIC255. Purple line: 4A11/PIC255.

FIG. 7. ME voltages when a 3A peak triangle current (1kHz) flows in the 8 turns coil (CRA configuration). Dotted line: TFD/PIC255 sample. Dashed line: FNCZ/PIC255 sample. Solid line: voltage obtained with a commercial active current probe (sensitivity: 0.5 V/A).

FIG. 8. ME voltages when a 300mA peak triangle waveform current (1kHz) flows in the 8 turns coil wounded on the FNCZ/PIC255 sample (CRA configuration). Thin solid line: voltage obtained with a commercial active current probe (sensitivity: 0.5 V/A). Dotted line : voltage *V* on a half electrode. Dashed line: voltage *V'* on the other half electrode. Thick solid line: differential voltage between the two electrodes.

FIG. 9. 3E8/PIC255 sample: fundamental (circles), second-harmonic (squares), and third-harmonic (triangles) of ME voltages versus AC field (1kHz) amplitude (CRA configuration). The thin lines correspond to voltages measured on a half electrode. The thick lines correspond to voltages measured between the two halve electrodes (differential voltages).

FIG. 10. FNCZ/PIC255 sample: fundamental (circles), second-harmonic (squares), and third-harmonic (triangles) of ME voltages versus AC field (1kHz) amplitude (CRA configuration). The thin lines correspond to voltages measured on a half electrode. The thick lines correspond to voltages measured between the two halve electrodes (differential voltages).

FIG. 11. TFD/PIC255 sample: fundamental (circles), second-harmonic (squares), and third-harmonic (triangles) of ME voltages versus AC field (1kHz) amplitude (CRA configuration). The thin lines correspond to voltages measured on a half electrode. The thick lines correspond to voltages measured between the two halve electrodes (differential voltages).

FIG. 12. 3E8/PIC255 sample: fundamental (solid line) and second-harmonic (dotted line) ME voltages in CCA configuration. The ampere-turns are produced by a 30 turns coil wounded on the ME ring and carrying a DC current.

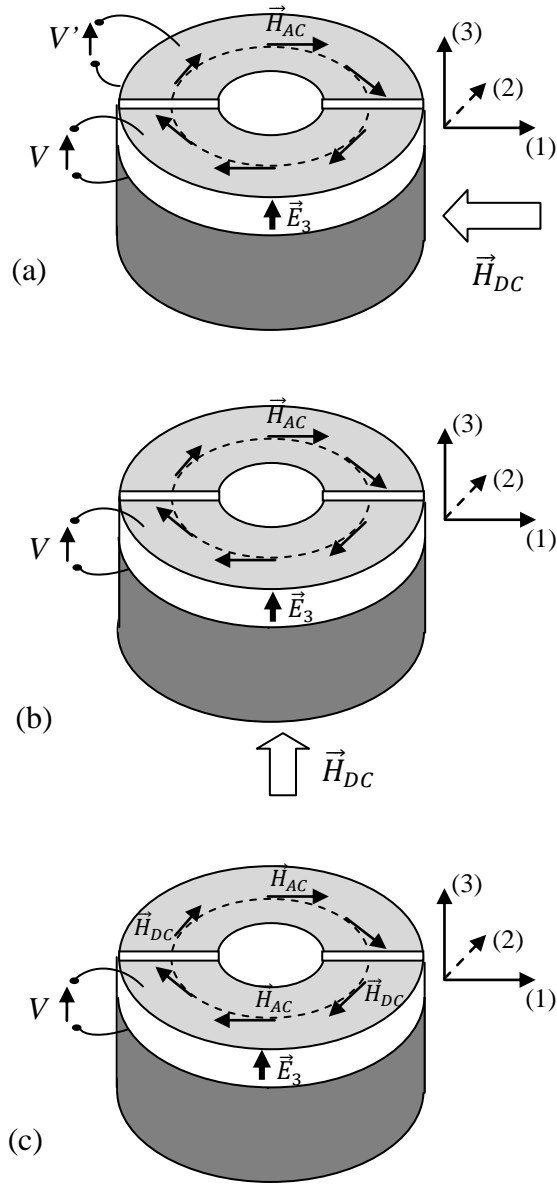

FIG.1. Sketches of three configurations of ME measurement. (a): the circumferential AC magnetic field is forced in the (1,2) plane (the coil is not represented) and the DC magnetic field is applied in the (radial) direction (1) (CRA). (b): the circumferential AC magnetic field is forced in the (1,2) plane and the DC magnetic field is applied in the (axial) direction (3) (CAA). (c): both AC and DC magnetic fields are forced in the circumferential direction (CCA). In all case, the electric field is measured in the (axial) direction (3).

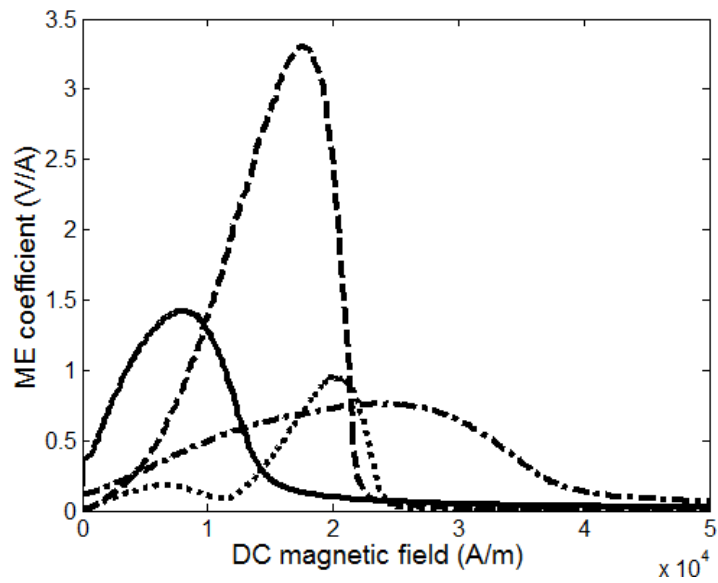

FIG. 2. Transversal magnetoelectric coefficients under radial DC field (CRA configuration). Dashed line: 3E8/PIC255 sample. Dotted line: 3E6/PIC255 sample. Solid line: 4A11/PIC255 sample. Dashed-dotted line: FerriteNiCoZn/PIC255.

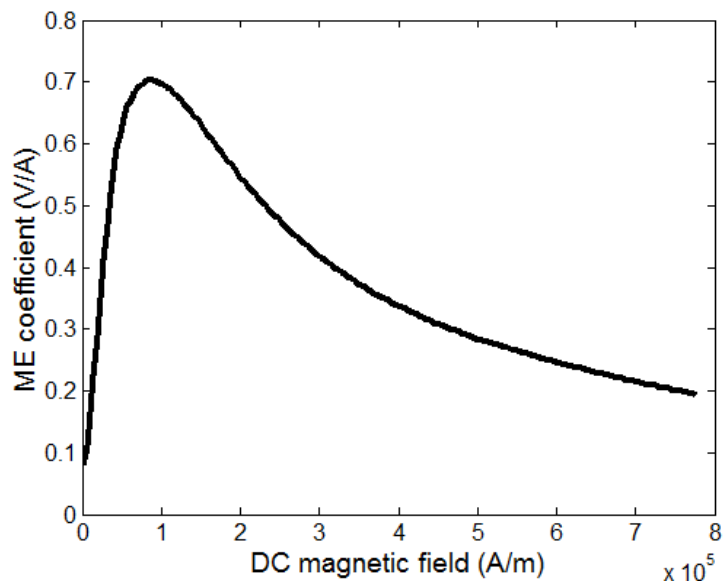

FIG. 3. Transversal magnetoelectric coefficient under radial DC field (CRA configuration) for the TFD/Pic255 sample.

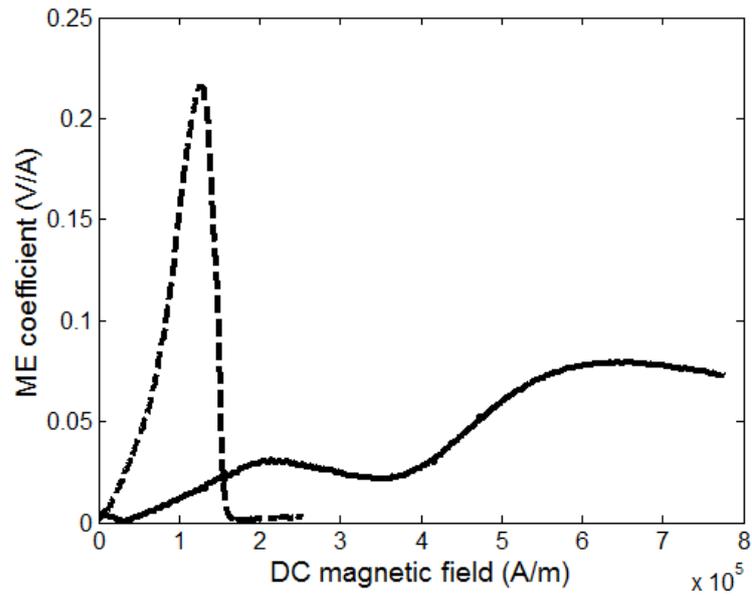

FIG. 4. Magnetoelectric coefficients when the DC field is applied in (axial) direction (3) and the AC field is in the (1,2) plane (CAA configuration). Dashed line: 3E8/Pic255 sample. Solid line: TFD/Pic255 sample.

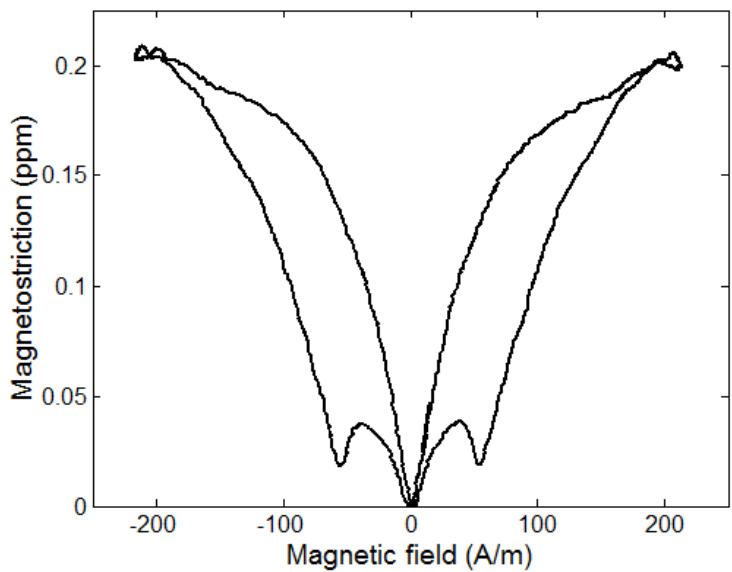

FIG. 5. Longitudinal magnetostristion curve versus internal magnetic field measured on a 3E6 ring excited at 1kHz.

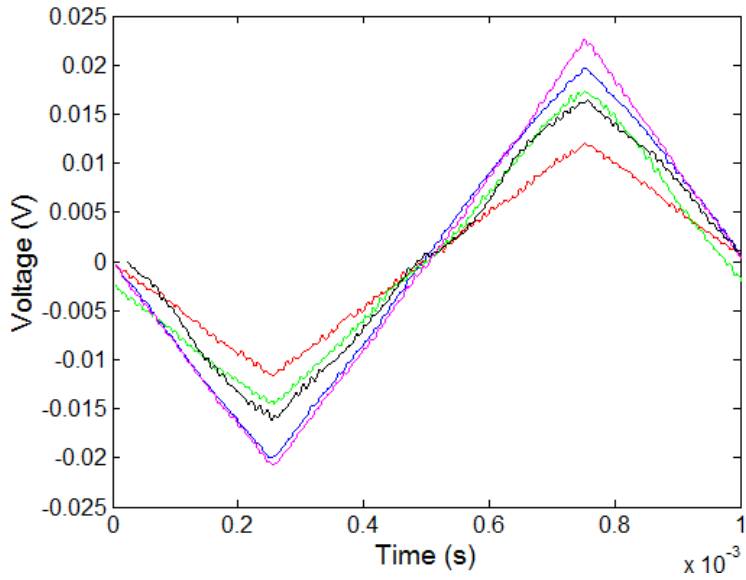

FIG. 6. ME voltages when a 55mA peak triangle current flows in the 8 turns coil (CRA configuration). Red line: TFD/PIC255 sample. Black line: FNCZ/PIC255 sample. Green line: 3E6/PIC255 sample. Blue line: 3E8/PIC255. Purple line: 4A11/PIC255.

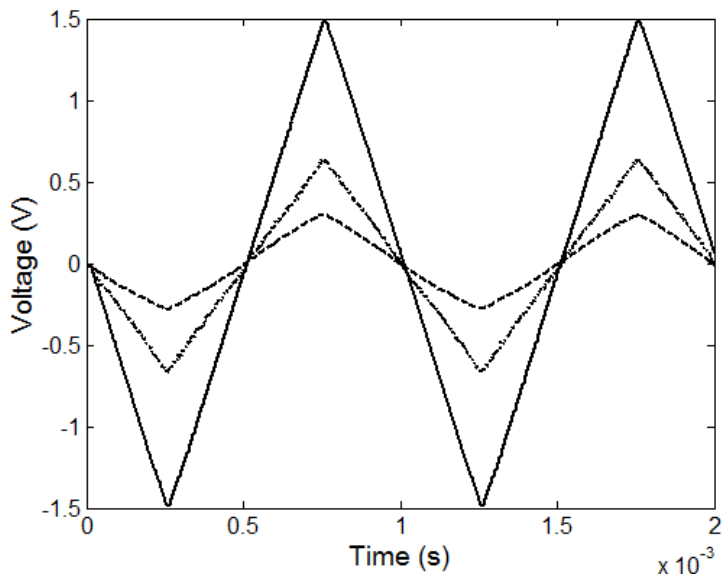

FIG. 7. ME voltages when a 3A peak triangle current (1kHz) flows in the 8 turns coil (CRA configuration). Dotted line: TFD/PIC255 sample. Dashed line: FNCZ/PIC255 sample. Solid line: voltage obtained with a commercial active current probe (sensitivity: 0.5 V/A).

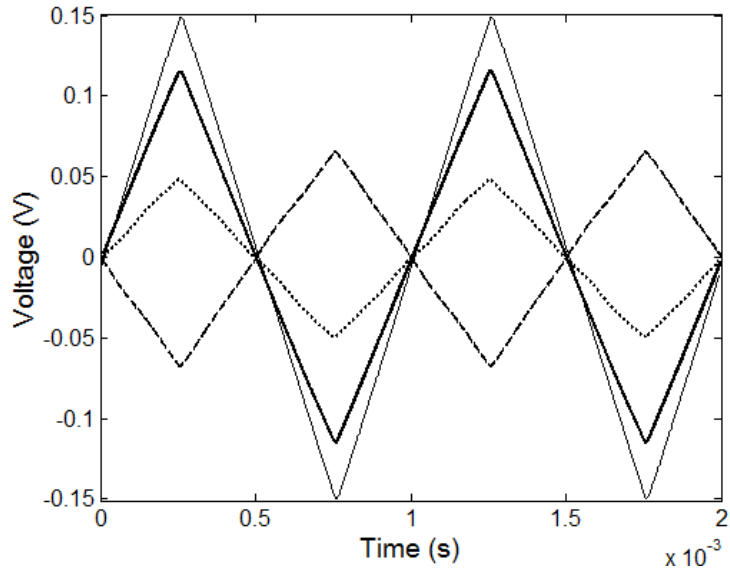

FIG. 8. ME voltages when a 300mA peak triangle waveform current (1kHz) flows in the 8 turns coil wounded on the FNCZ/PIC255 sample (CRA configuration). Thin solid line: voltage obtained with a commercial active current probe (sensitivity: 0.5 V/A). Dotted line : voltage $V$ on a half electrode. Dashed line: voltage $V'$ on the other half electrode. Thick solid line: differential voltage between the two electrodes.

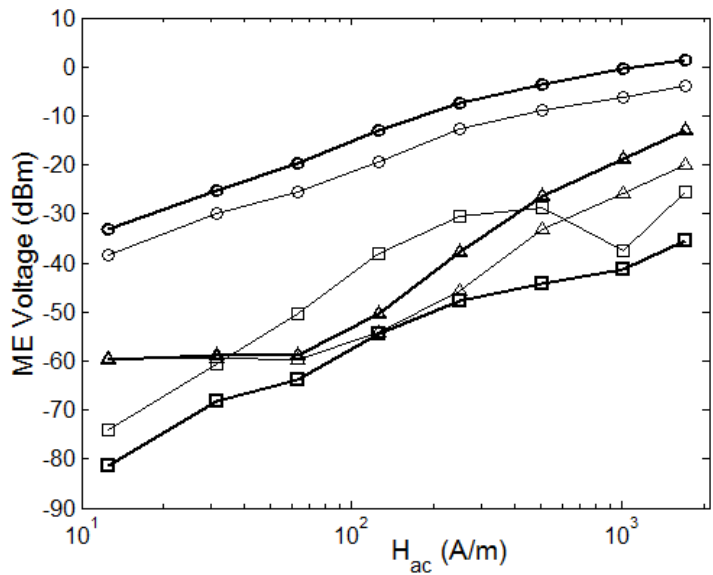

FIG. 9. 3E8/PIC255 sample: fundamental (circles), second-harmonic (squares), and third-harmonic (triangles) of ME voltages versus AC field (1kHz) amplitude (CRA configuration). The thin lines correspond to voltages measured on a half electrode. The thick lines correspond to voltages measured between the two halve electrodes (differential voltages).

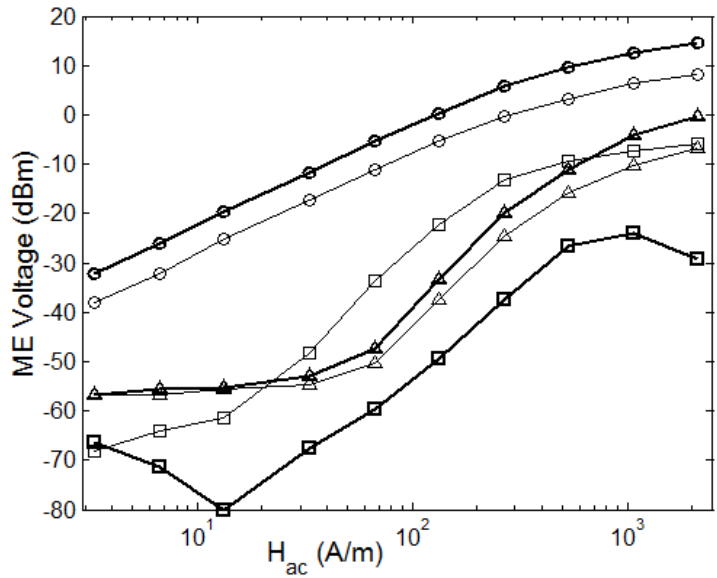

FIG. 10. FNCZ/PIC255 sample: fundamental (circles), second-harmonic (squares), and third-harmonic (triangles) of ME voltages versus AC field (1kHz) amplitude (CRA configuration). The thin lines correspond to voltages measured on a half electrode. The thick lines correspond to voltages measured between the two halve electrodes (differential voltages).

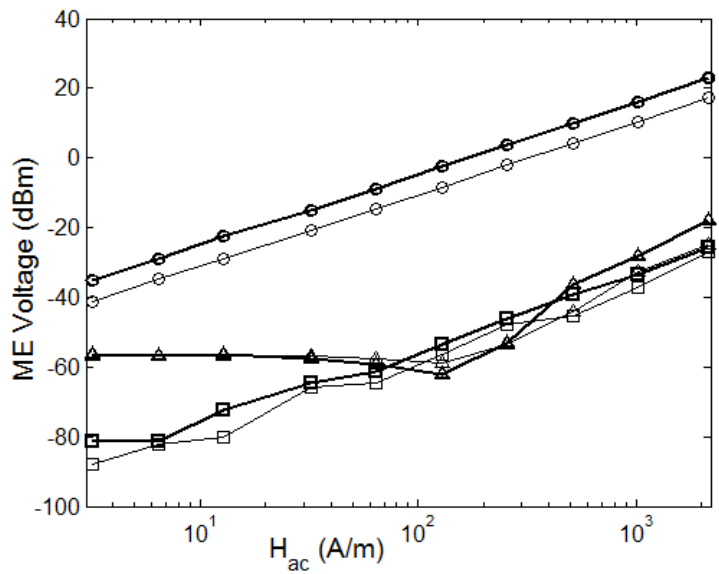

FIG. 11. TFD/PIC255 sample: fundamental (circles), second-harmonic (squares), and third-harmonic (triangles) of ME voltages versus AC field (1kHz) amplitude (CRA configuration). The thin lines correspond to voltages measured on a half electrode. The thick lines correspond to voltages measured between the two halve electrodes (differential voltages).

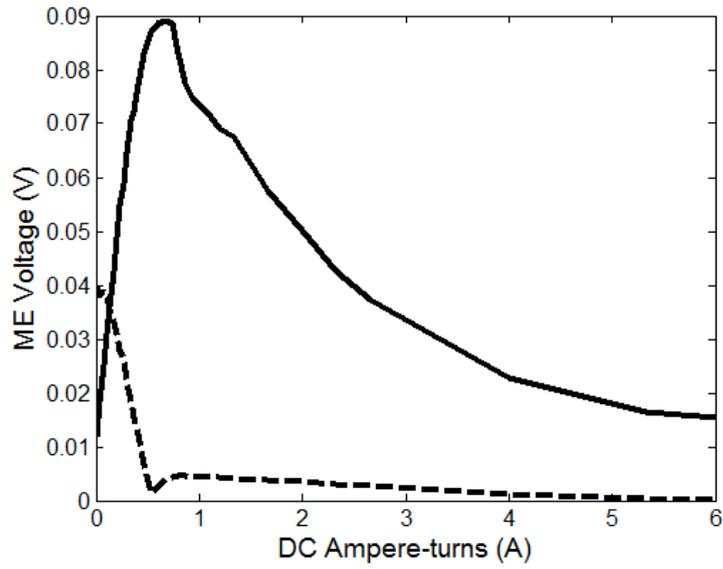

FIG.12. 3E8/PIC255 sample: fundamental (solid line) and second-harmonic (dotted line) ME voltages in CCA configuration. The ampere-turns are produced by a 30 turns coil wounded on the ME ring and carrying a DC current